\documentclass[aps,article,nofootinbib,twocolumn]{revtex4}

\usepackage{epsfig}
\usepackage{graphicx}
\usepackage{graphics}

\usepackage{amsfonts}  
\usepackage{amsmath, amssymb, stmaryrd}     

\usepackage{afterpage}  
\usepackage[usenames, dvipsnames]{color}
\definecolor{ao(english)}{rgb}{0.93, 0.53, 0.18}

\def\be{\begin{equation}}
\def\ee{\end{equation}}
\def\bi{\begin{itemize}}
\def\ei{\end{itemize}}

\begin{document}
\title{Detecting  Domain Walls in Laboratory Experiments}
\author{Claudio Llinares}
\affiliation{Institute for Computational Cosmology, Department of Physics, Durham University, Durham DH1 3LE, U.K.}

\author{Philippe Brax}
\affiliation{Institut de Physique Th\'eorique, Universit\'e Paris-Saclay, CEA, CNRS, F-91191 Gif-sur-Yvette Cedex, France}

\begin{abstract}
The inherently unstable nature of domain walls makes their detection in laboratory experiments extremely challenging.  We propose a method to stabilise domain walls in a particular modified gravity model inside a cavity.  We suggest two ways in which the walls could  be detected once stabilized: studying the trajectories of Ultra Cold Neutrons (UCN's) either via  the deflection angle of a neutron beam induced by  the attraction towards the wall or through the time difference of these particles passing through the wall.  We give realistic estimates for these effects and expect that they should be detectable experimentally.
\end{abstract}

\maketitle

\section{Introduction}

Domain walls are a particular class of topological defects that exist in many areas of  physics \citep{2000csot.book.....V, 2006kdwi.book.....V}.  Here we deal with domain walls which may appear in the gravity sector of alternative theories for gravity, where the scalar field whose profile is responsible for the existence of the topological defect couples to matter in a crucial way.  There is a large spectrum  of theoretical works studying the properties of topological defects in cosmology using both  analytical \citep{1996PhRvL..77.4495H, 2003PhRvD..68d3510H, 1984PhRvD..30.2036V} and numerical techniques \citep{1989ApJ...347..590P, 1990PhRvD..41.1013K, 1996PhRvD..53.4237C, 1997PhRvD..55.5129L, 2003PhRvD..68j3506G, 2005PhRvD..72h3506A, 2011PhRvD..84j3523L, 2013PhLB..718..740L}.  However, efforts to detect them with cosmological observations have failed so far \citep{2014A&A...571A..25P}.

This is the reason why we focus on the detection in laboratory experiments. Unfortunately it turns out that one has to face the fact that  domain walls are unstable. As a result, even if it may be possible to form them in a cavity, they will not last long enough to be detected with current experiments \citep{2015JCAP...03..042B, 2016ConPh..57..164B, 2016JCAP...08..070B, 2016JCAP...12..041B}.  The reason for this instability in the case for  which the walls are coupled to matter is that the defects minimize the energy in regions where the matter density is high, thus they tend to move towards the walls of the experiment and disappear altogether \citep{2014PhRvD..90l4041L, 2016JCAP...12..041B, 2017PhRvD..95f4047P}.  We show here that the coupling to matter, which generates the aforementioned issue  in the first place, leads to its  actual  solution:  A domain wall can be stabilized inside a cavity by introducing a distribution of matter in the centre of the experiment, where the wall can attach itself.  We describe this mechanism in detail in a realistic set up.  Furthermore, we propose possible strategies for detecting these walls once they are stabilized inside the cavity.  This involves the monitoring of neutral particle trajectories such as UCN's across the wall or grazing the wall. In both cases, either via the time difference compared to the situation with no wall or the deflection angle induced by the attraction towards the wall, we find that for particle beams with macroscopic velocities in the m/sec ballpark, and symmetron parameters compatible with previously studied experimental situations\cite{Jaffe:2016fsh,Brax:2016wjk}, such as atomic interferometry,  the resulting effects should be detectable.

\section{The symmetron model}

The technique that we will use to stabilize the domain walls in the cavity of the experiment requires the scalar field to be coupled to matter.  In order to obtain predictions for a realistic set up, we work with the symmetron model, which was first discussed in \cite{2010PhRvL.104w1301H}.  Predictions and constraints to the model exist in several contexts and spanning several orders of magnitude in scales including laboratory experiments \citep{2013PhRvL.110c1301U, 2016JCAP...12..041B}, Solar system scales \cite{2010PhRvL.104w1301H, 2017PhRvL.118j1301H}, galaxy scales \citep{2017PhRvD..95f4050B, 2018MNRAS.476L..29L, 2018arXiv180505226O, 2012JCAP...01..030C}, galaxy clusters \cite{2014A&A...562A...9G, 2013PhRvL.110o1104L, 2014A&A...562A..78L, 2015MNRAS.449.2837G}, cosmological scales including large scale structure of the Universe \citep{2011PhRvD..84j3521H, 2011PhRvD..84l3524B, 2017MNRAS.472L..80L, 2016A&A...595A..40I, 2014PhRvD..89b3523T} and variations of fundamental constants \citep{2014PhRvD..89b4025S, 2017PhLB..769..491P}.  Furthermore, several N-body cosmological codes can simulate the model \cite{2012ApJ...748...61D, 2012JCAP...10..002B, 2014A&A...562A..78L, 2013PhRvL.110p1101L, 2014PhRvD..89h4023L, 2015MNRAS.454.4208W, 2016A&A...585A..37H}.

%
\begin{figure*}
  \begin{center}
    \includegraphics[width=\textwidth]{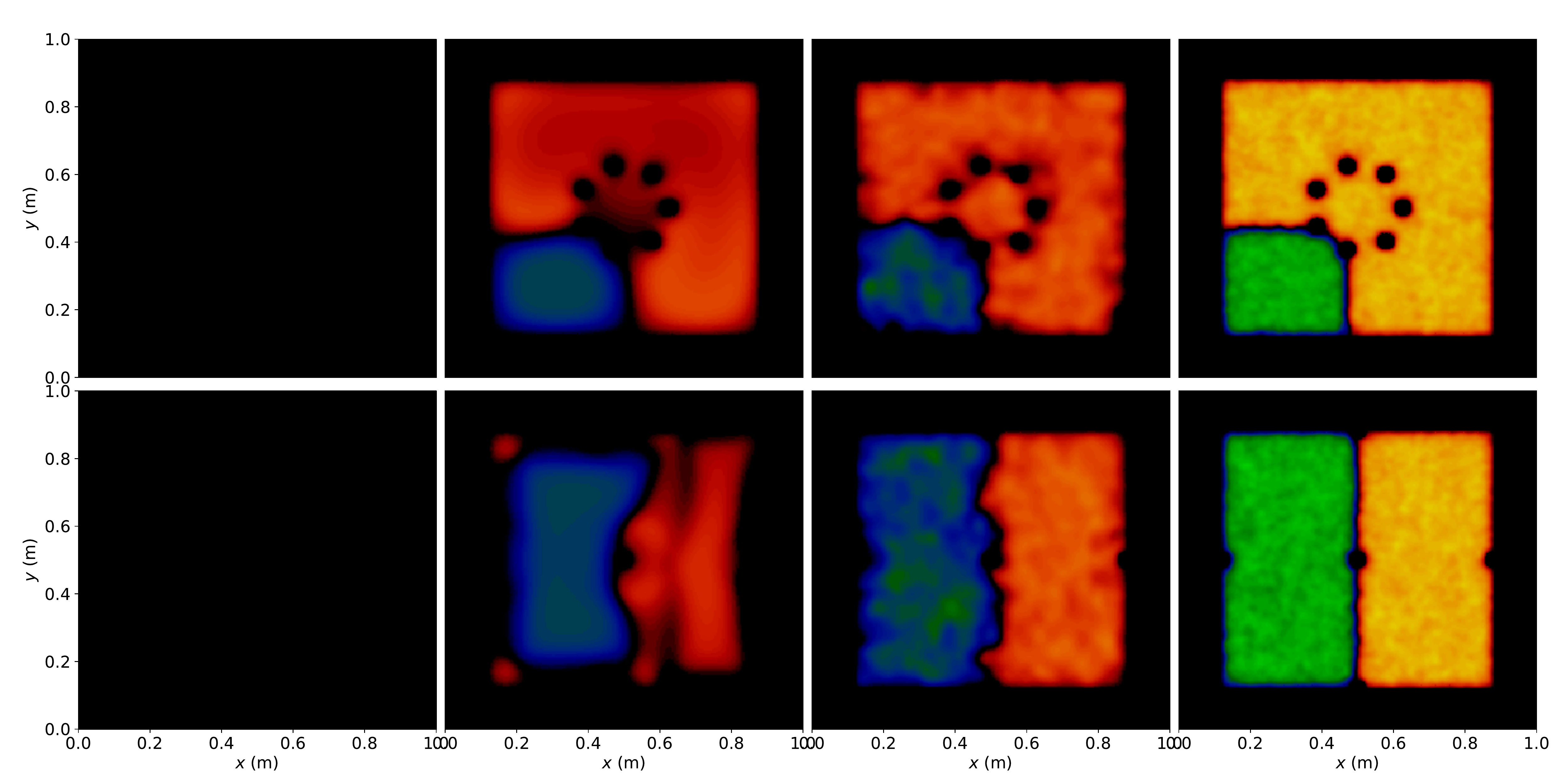}
    \caption{Four stages of the generation of stable domain walls in a cavity for two different configuration of stabilizers.  The colours correspond to the scalar field normalized to its vacuum value. The black regions around each panel correspond to the walls of the cavity, where the scalar field remains screened during the whole process.}
    \label{fig:four_panels}
  \end{center}
\end{figure*}

The model is defined with the following action
\be
S = \int \sqrt{-g} \left[ R - \frac{1}{2}\nabla^a\phi \nabla_a \phi - V(\phi)\right] d^4x + S_M(\tilde{g}_{ab}, \psi) \ ,
\label{symm-action}
\ee
where the Einstein $g_{ab}$ and the Jordan $\tilde{g}_{ab}$ frame metrics are conformally related (i.e. $\tilde{g}_{ab} = A^2(\phi) g_{ab}$).  The potential and the conformal factor have the following forms:
\begin{align}
\label{potential}
V(\phi) &= -\frac{1}{2}\mu^2\phi^2 + \frac{1}{4}\lambda\phi^4 + V_0 \\
\label{conformal_factor}
A(\phi) &= 1 + \frac{1}{2}\left(\frac{\phi}{M}\right)^2 \ ,
\end{align}
where $\mu$ and $M$ are mass scales which define the nature of the effective potential in the presence of matter, and  $\lambda$ is a dimensionless constant. When coupled to non-relativistic matter like the walls of a cavity in a laboratory experiment,  the equation of motion of the symmetron field takes the following form when assuming a Minkowski background, as always realistic in an earth-bound experimental context, and the quasi-static approximation for matter sources:
\be
\nabla^2\phi = \left( \frac{\rho}{M^2} - \mu^2 \right)\phi + \lambda\phi^3.
\label{field_eq}
\ee
The validity of this approximation in a cosmological context was tested in Refs.~\citep{2014PhRvD..89h4023L, 2015MNRAS.454.4208W}.  In a nutshell  the scalar field evolves  in the following effective potential
\be
V(\phi)_{\mathrm{eff}} = \frac{1}{2}\left(\frac{\rho}{M^2} - \mu^2\right)\phi^2 + \frac{1}{4}\lambda\phi^4 + V_0.
\label{def_effective_potential}
\ee
Notice that the potential is $Z_2$ symmetric and that symmetry breaking occurs at low energy. Indeed in regions of space when the density is smaller than $\rho_{SSB} = \mu^2 M^2$, the potential has two minima and thus, can give rise to domain walls.  In the opposite case, the  $Z_2$ symmetry of the model is restored and the scalar field is forced to oscillate around zero, screening the effects of its associated fifth force and allowing it to be consistent with Solar System observations \citep{2010PhRvL.104w1301H}.  This density dependence is also responsible for the unique properties of the domain walls, which were studied in depth in \citep{2014PhRvD..90l4041L, 2017PhRvD..95f4047P}.  In particular, the coupling to matter forces the walls to attach themselves to high density regions, which is a well-know phenomenon in the context of ferromagnetism, where domain walls interact with impurities (see e.g.~\cite{jiles_hysteresis} and references therein).  The following section describes how this effect can be used to stabilize a wall inside vacuum cavities that are employed in usual experimental setups.

\section{Stabilization of a coupled domain wall inside a cavity}

As we know that coupled domain walls are attracted to high density regions, we propose to stabilize them by introducing high density areas inside the vacuum chamber of the experiment.  Thus the process by which domain walls can be stabilized is the following:  a phase transition is forced by generating vacuum inside the cavity, which originally has to be filled with a gas of density higher than $\rho_{SSB}$.  As the density is reduced and during the resulting phase transition, the field can fall into one or the other minima of the effective potential and thus domain walls are formed. We propose to realise this scenario in the laboratory where the cooling process and the phase transition which could have happened in the Universe for cosmological symmetrons when the mass parameter $\mu$ is taken to be of relevance cosmologically (i.e. a small value of $\mu$ such that the range of the scalar induced force is a fraction of the cosmological horizon now take place inside a cavity).  We shall focus on values of $\mu$ which are much larger and adapted to the modest size of experimental cavities (i.e. for the range $\mu^{-1}$ of the scalar interaction smaller than the size of the cavity). The resulting  walls inside the cavity will evolve following their own equations of motion, searching for high density regions where they can minimize their energy.  If any of the created walls happen to pass by the stabilising element and the geometry of the stabilizer is correctly  chosen to be adapted  to the model parameters, there will be  a non-negligible probability that it will stay attached to it and become stable.

We tested this process with 2D simulations which were run with a 2D version of the non-static code presented in \citep{2013PhRvL.110p1101L}, which was also used to study properties of symmetron domain walls in \citep{2014PhRvD..90l4041L}.  As the simulations are 2D, the stabilizers are filaments, for which we choose a circular section.  Additional filaments lying in the walls of the cavity may be added to gain more control on the geometry of the final configuration.  In these particular examples, the cavity is a 2D box with a size of 1 meter.

Figure \ref{fig:four_panels} shows the four stages of the experiment for two different configurations of the stabilizers:
\bi
\item \textit{Left: Initial condition.}
The cavity is filled with gas and thus, the scalar field is screened, i.e. its average value vanishes,   with minimal perturbations around zero.
\item \textit{Middle left:  Symmetry breaking and formation of a wall.}
Once the density of the gas inside the cavity falls below the density of symmetry breaking $\rho_\mathrm{{SSB}}$, the scalar field collapses to one or the other minima of the effective potential in different regions of space.  At this stage, there is no guarantee that the resulting domain walls will survive.
\item \textit{Middle right:  A survival domain wall is trapped by the filaments.}
The wall is now stable and will not collapse towards the walls of the cavity.  Strong perturbations still exist in the wall, which does not have a straight configuration.
\item \textit{Right:  Domain wall fully stabilized and ready to be detected.}
The perturbations in the wall are transferred to a background of scalar waves that travel through the whole cavity and are reflected on its sides.  By doing this, the domain wall loses its kinetic energy and adopts an almost straight configuration, with minimal perturbations.
\ei
The configuration of the filaments are the simplest ones for which we found stable domain walls.  However, this configuration is not unique.  Different configurations of filaments may affect the way the phase transition occurs and increase the chances of forming a wall when releasing the gas.  Furthermore, moving sheets of material can be used to control the initial configuration of the walls after the phase transition happens by forcing a particular geometry.  Some details on how the geometry affects the stability of the walls was presented already in \citep{2014PhRvD..90l4041L}.

\section{Detection of a domain wall inside a cavity}

Once a domain wall is formed and stabilized inside the vacuum chamber of the experiment, we need to find a way to detect it.  We propose here two different setups, which are based on how the domain wall affects the trajectories of slow particles such as UCN's as  presently used to search for chameleon fields \cite{Lemmel:2015kwa,Brax:2013cfa,Brax:2011hb,Jenke:2011zz}.  These neutrons can have energies of the order of $10^{-9} ~ \mathrm{eV}$, which correspond to velocities of the order of 1 m/sec, which is the velocity we assume for the neutrons.  Furthermore, in order to obtain realistic estimates for detectability, we  assume values for the free parameters of the symmetron field which are compatible with the ones already probed by atomic interferometry technique:
\be
(\mu_0, M_0, \lambda_0) = (2.4\times10^{-3} ~ \mathrm{eV}, 10^{9} ~ \mathrm{eV}, 0.1).
\label{fiducial_model}
\ee
 Notice that the value of $\mu$ is taken to be of the order of the dark energy scale. This is compatible with that fact that the quantum fluctuations due to the symmetron field of order $\mu^4$ could be at the origin of the acceleration of the expansion of the Universe. On the other hand, this large value of $\mu$ compared to the Hubble rate now $H_0$ implies that the scalar interaction mediated by the symmetron would have such a short range in vacuum, of the order of $0.1$ mm, that no fifth force manifestation of the symmetron would occur in the large scale structure of the Universe.

\begin{figure}
  \begin{center}
    \includegraphics{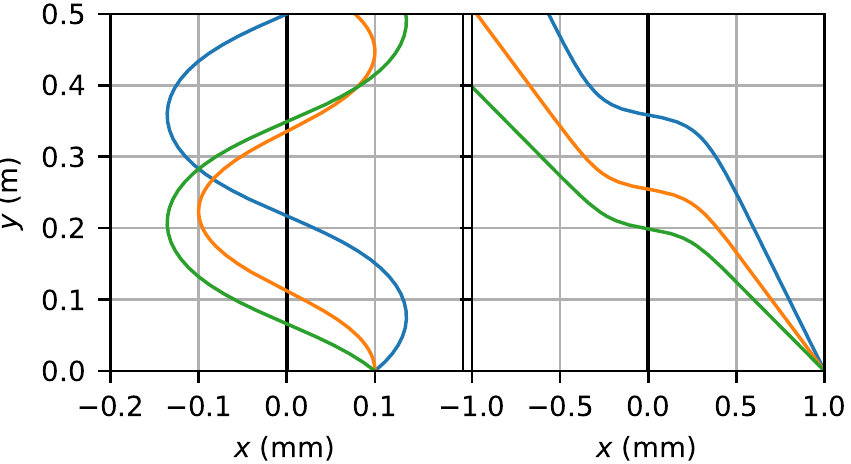}
    \caption{Examples of trajectories of UCN's in the surroundings of a wall.  The trajectories to the left and right correspond to values of $a$ greater and smaller than one.  Different curves correspond to different angles of the initial velocity.  Note that the horizontal axis have different scale in different panels.}
    \label{fig:some_trajectories}
  \end{center}
\end{figure}

\subsection{Trajectory of a massive particle around a domain wall}

In order to determine the trajectory of the neutrons in the vicinity  of a domain wall, we need to solve Newton's equations, which in the symmetron case are given by:
\be
\ddot{\textbf{x}} = -c^2 \frac{\nabla\left(\phi^2\right)}{2M^2}.
\label{eq_particles}
\ee
The effects of the gravitational field of the Earth can be effaced by choosing the filaments to be vertical and studying the motion of particles in the horizontal plane.  The solution in the $x$ direction, which we chose as perpendicular to the wall and parallel to the earth surface, can be obtained by taking into account that the Hamiltonian of the particle is conserved in each direction separately.  So we can replace Newton's equation in the $x$ direction with the conservation equation
\be
\label{Hx}
H_x = \frac{\dot{x}^2}{2} + \frac{c^2\phi^2}{2M^2} = \mathrm{constant}.
\ee
Using the domain wall solution  \citep{2000csot.book.....V, 2006kdwi.book.....V} of the field equation (Eq.~\ref{field_eq}) and integrating once, we obtain the constraint
\be
\int^{x(t)}_{x_0} \frac{dx}{\sqrt{1-a \tanh^2\left(b x\right)}} = \sqrt{2H_x} (t-t_0),
\ee
where we have used the following definitions:
\begin{align}
\label{def_a}
a & \equiv \frac{c^2\phi_0^2}{2H_x M^2}, &
b & \equiv \frac{1}{2\lambda_0}, \\
\phi_0 & = \frac{\mu}{\sqrt{\lambda}}, &
\lambda_0 & = \frac{\hbar c}{\sqrt{2}\mu},
\end{align}
where $\phi_0= \frac{\mu}{\sqrt\lambda}$ is the vacuum expectation value of the field and $\lambda_0= 1/m_0$ is its Compton wavelength in vacuum (see Ref.~\citep{2014PhRvD..90l4041L} for non-vacuum solutions).  The numerical value of $\phi_0$ for our fiducial model defined by Eq.~\ref{fiducial_model} is $7.5\times10^{-3}~\mathrm{eV}$.  The integrals can be done analytically and  we obtain the following solution for the trajectory of a test particle:
\be
x(t) = \begin{cases}
\frac{1}{b} \mathrm{asinh}\left[ \frac{\sinh\left( \omega t + \alpha \right)}{\sqrt{1-a}}  \right] & \text{if $a<1$},\\
\frac{1}{b} \mathrm{asinh}\left[ \frac{\sin\left( \omega t + \alpha \right)}{\sqrt{a-1}} \right] & \text{if $a>1$}
\end{cases}
\ee
\begin{figure}[!t]
  \begin{center}
    \includegraphics{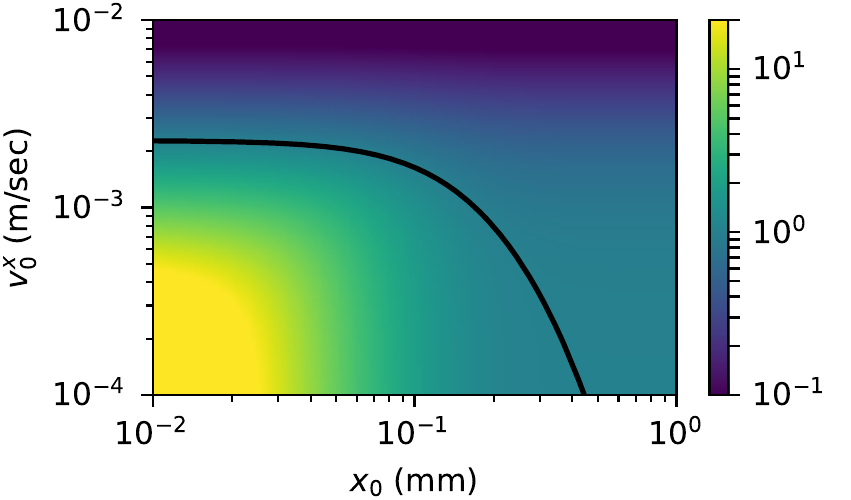}
    \caption{Values of constant $a$ define by Eq.\ref{def_a} as a function of the initial position and velocity of particle with respect to a domain wall.}
    \label{fig:a_of_x0_v0}
  \end{center}
\end{figure}
and its inverse:
\be
t(x) = \begin{cases}
\frac{1}{\omega} \left\{ \mathrm{arcsinh}\left[\sqrt{1-a}\sinh\left( b x \right)\right] - \alpha \right\} & \text{if $a<1$},\\
\frac{1}{\omega} \left\{ \arcsin\left[\sqrt{a-1}\sinh\left( b x \right) \right] - \alpha \right\} & \text{if $a>1$}.
\end{cases}
\ee
Notice that there are two types of solutions depending on $a$, i.e. depending on the energy of the neutron in the $x$ direction.  Figure \ref{fig:some_trajectories} shows typical trajectories for initial conditions associated with values of $a$ greater (left) and smaller (right) than one.  At low energy, the neutron oscillates around the domain wall due to the scalar attraction whilst at higher energy the neutron is refracted with a deflection depending on the strength of the scalar interaction. The characteristic energy separating the two types of behaviour is given by $E_x= m_N \frac{\mu^2}{2 \lambda M^2}$ where $m_N$ is the neutron mass. The solutions are governed by the frequencies and initial phases
\be
\omega = b\sqrt{2H_x |a-1|} = b v_0^x
\ee
and
\be
\alpha = \begin{cases}
\mathrm{arcsinh} \left[ \sqrt{1-a}\sinh(b x_0)  \right] & \text{if $a<1$},\\
\mathrm{arcsin} \left[ \sqrt{a-1}\sinh(b x_0)  \right] & \text{if $a>1$}
\end{cases}
\ee
respectively.  The dependence of $a$ with the initial conditions is shown in Figure \ref{fig:a_of_x0_v0}.  The contour in this figure describes the set of initial conditions for which $a=1$ (i.e. for the energy in the $x$ direction equal to $E_x$).   Solutions are oscillatory for particles that are close enough to the wall and have a small transverse velocity.

\begin{figure}[!t]
  \begin{center}
    \includegraphics{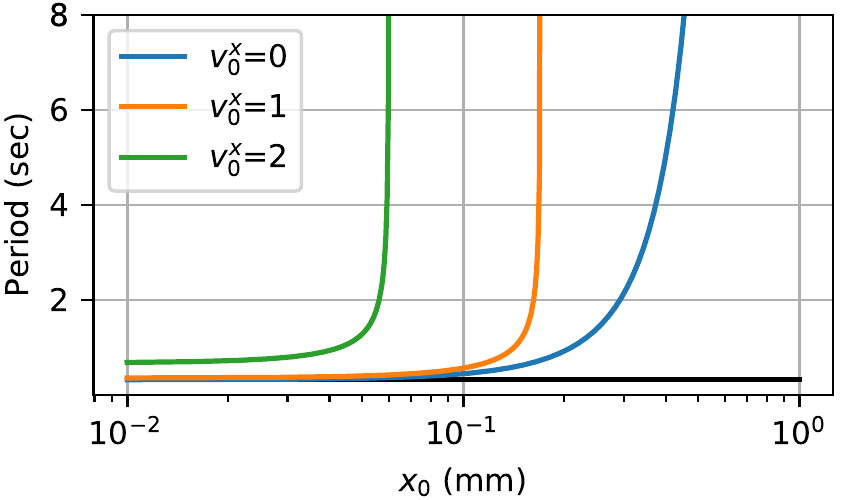}
    \caption{Period of oscillations of a particle moving around a wall as a function of its initial position with respect to the wall.  The three curves correspond to three different transverse velocities in m/sec.  The horizontal line is the minimum period given in Eq.\ref{pmin}.}
    \label{fig:period_of_x0}
  \end{center}
\end{figure}

\subsection{Proposed experiments}

The solution to the motion of a neutral and test particle (e.g.~a neutron) provided in the previous section shows that it makes sense to consider two kinds of experiments.  In the first one, neutrons can be launched for instance from the bottom wall of the vacuum chamber moving upwards (i.e. with $\textbf{v} = (\epsilon,1)$ m/sec).  Small values of $\epsilon$ will give oscillatory trajectories.  The period of these oscillations around the wall is shown in Figure \ref{fig:period_of_x0} for different values of the transverse velocity $\epsilon$.  Since the periods are of the order of a few seconds and the total velocity of the neutrons is of the order of a few m/sec, the trajectory of the neutrons will have a macroscopic displacement with respect to the trajectory that they would  have in the absence of the domain wall and thus, could be detected.  Note that there  exists a minimal period with corresponds to the limit $(x_0, v^x_0)\rightarrow (0,0)$:
\be
P_\mathrm{min} = 2\pi \frac{\sqrt{2\lambda}M \hbar}{\mu^2}.
\label{pmin}
\ee
The numerical value of $P_\mathrm{min}$ for the fiducial model defined by Eq.~\ref{fiducial_model} is 0.3 seconds.

\begin{figure}
  \begin{center}
    \includegraphics{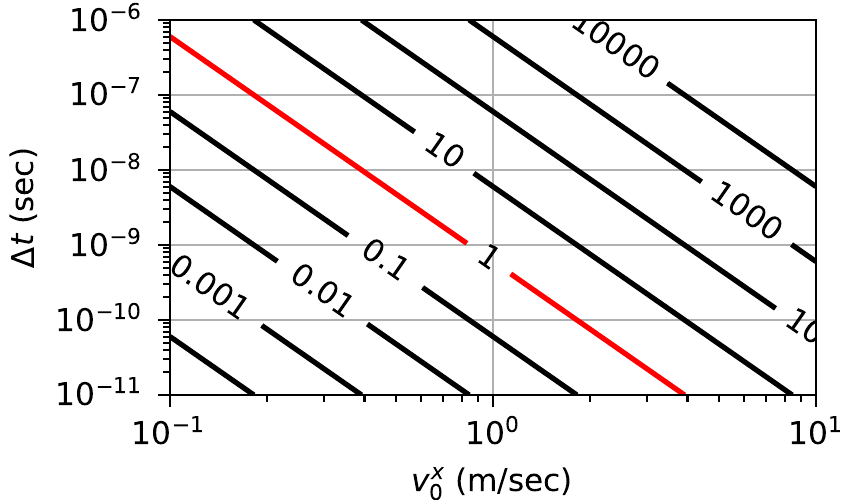}
    \caption{Difference in arrival time of UCN's that travel across a wall in a direction perpendicular to it as a function of the initial velocity with respect to the wall and for different multiples of $\eta_0$ associated to the fiducial model defined in Eq.~\ref{fiducial_model}.}
    \label{fig:delta_t_of_vx}
  \end{center}
\end{figure}

A different experiment could  be realised by launching the neutrons in the transverse direction, for instance, moving from the left to the right walls of the cavity (i.e. with $\textbf{v} = (1,0)$ m/sec).  These initial conditions are in the regime $a<1$ and thus, the particles will move through the wall having only a small perturbation in their transverse velocity:  particles will accelerate when approaching the wall and decelerate when moving away from it.  Thus, there will be a difference in the arrival times between perturbed and unperturbed trajectories.  This difference can be calculated by analysing the asymptotic behaviour of the trajectory in the initial and final positions of the particle:
\be
\Delta t = \sqrt{2} \eta  \frac{\hbar c^3}{\left(v^x_0\right)^3},
\ee
where we have condensed the dependence with the model parameters in the parameter $\eta = \mu/(M^2\lambda)$.  Typical values of $\Delta t$ are shown in Figure \ref{fig:delta_t_of_vx} for different multiples of the parameter $\eta$ associated to our fiducial model (Eq.~\ref{fiducial_model}).  This particular model in connection with UCN's travelling at 1 m/sec (red curve) gives a difference in arrival times of the order of $10^{-9}$ sec.

\bigskip
\section{Conclusions}

We have described the stabilisation of domain walls in the case of symmetrons in a cavity when stabilising filaments are present. The walls are the result of the symmetry breaking phase transition in the cavity when the gas density is lowered below the symmetry breaking scale. We have explicitly shown numerically how this process can be realised in an efficient way. Moreover we have illustrated the presence of the walls in a cavity by calculating the measurable effects that such wall would have on Ultra Cold Neutron trajectories. It turns out that for neutrons with velocities around $1$ m/s the deviation of their trajectories would be macroscopic for symmetron models with a scalar force range of the order of $0.1$ mm as previously tested by atomic interference experiments and whose quantum fluctuations would be in the right ball park to generate the late time acceleration of the expansion of the Universe.

\acknowledgments CLL acknowledges support from STFC consolidated grant ST/L00075X/1 \& ST/P000541/1 and ERC grant ERC-StG-716532-PUNCA.  This work is supported in part by the EU Horizon 2020 research and innovation programme under the Marie-Sklodowska grant No. 690575. This article is based upon work related to the COST Action CA15117 (CANTATA) supported by COST (European Cooperation in Science and Technology).

\bibliography{references}

\end{document}